# COST-BASED ASSESSMENT OF PARTITIONING ALGORITHMS OF AGENT-BASED SYSTEMS ON HYBRID CLOUD ENVIRONMENTS

**Chahrazed Labba[a], Narjès Bellamine Ben Saoud[a] [b]**

[a] Laboratoire RIADI-Ecole Nationale des Sciences de l'Informatique, Univ Manouba Tunisia.
[b] Institut Supérieur d'Informatique, Univ. Tunis El Manar, Tunisia.

[a]chahrazedlabba@gmail.com, [b]narjes.bellamine@ensi.rnu.tn.

**ABSTRACT**

Distributing agent-based simulators reveals many challenges while deploying them on a hybrid cloud infrastructure. In fact, a researcher's main motivations by running simulations on hybrid clouds, are reaching more scalable systems as well as reducing monetary costs. Indeed, hybrid cloud environment, despite providing scalability and effective control over proper data, requires an efficient deployment strategy combining both an efficient partitioning mechanism and cost savings. In this paper, we propose a cost deployment model dedicated to distributed agent-based simulation systems. This cost model, combining general performance partitioning criteria as well as monetary costs, is used to evaluate cluster and grid based partitioning algorithms on hybrid cloud environments. The first experimental results show that, for a given agent-based model, a good partitioning method used with the suitable hybrid cloud environment lead to an efficient and economic deployment.

## 1. INTRODUCTION

Agent-based systems are largely used in modeling and simulating various types of complex systems in many research areas such as sociology (Macy and Willer 2002), crisis and emergency management (Bellamine Ben Saoud Ben Mena Dugdale Pavard and Ben Ahmed 2006), biology (Kuan Rui-bin Hao-ran and Jun-qing 2011), natural disasters (Mustapha Mcheick and Mellouli 2013), etc.

To cope with realistic complex situations, large-scale simulations are required. However, considering the large number of the simulated entities and their complex behaviors, scalability becomes a challenging issue. Therefore, in addition to distribution and suitable partitioning methods, using powerful computational infrastructures should ensure better performance of such large-scale simulation environments. As far as we know unlimited computational resources are not in the reach of all organizations and budget constraints are a strong consideration in choosing a solution. Cloud computing paradigm offers new opportunities by delivering limitless resources on demand and on pay-as-you-go basis. In the current work we are interested in using the hybrid clouds. In fact, such environments consist in combining a local infrastructure and a public cloud offered by a third party. Hybrid cloud permits to overcome the challenge of scalability that can be induced by using only the private machines and minimize the costs of outsourcing the whole data and computing on a public cloud. Such environments were used with different distributed application types such as web applications (Kaviani Wohlstadter and Lea 2012), enterprise applications (Hajjat Sun Sung Maltz Rao Sripanidkulchai and Tawarmalani 2010), agent-based systems (Siddiqui Tahir Rehman Ali Rasooland Bloodsworth 2012).

However, to make an efficient use of such infrastructure, both good distribution strategy and effective cost deployment model should be taken into consideration. To the best of our knowledge, hybrid clouds have not been used to run distributed large-scale agent-based simulations. In the current work we are interested in proposing an effective cost deployment model that aims at measuring the suitability of the distribution of a large scale agent-based system on a hybrid cloud environment. Indeed an efficient deployment of a given distributed agent-based model consists of using a good partitioning method in terms of performance criteria (minimizing communication overhead, migration and global execution time) and a suitable public cloud provider (the cheapest is the best, to ensure costs savings) . We used the proposed cost deployment model to evaluate and compare the performance and the effectiveness of two agent-based partitioning algorithms on hybrid clouds environments. Among others, we have chosen to evaluate: 1) A cluster-based algorithm represented by the K-means clustering approach which consists in clustering the agents based on a pre-defined criteria then assigning them to different machines.2) A grid-based algorithm spotted by the static environment partitioning approach which consists of decomposing the space component (spatial environment) into multiple portions.

Each portion together with the set of agents residing in it are assigned to a node.

For the hybrid cloud environments, we used two public cloud providers including Amazon Elastic Compute Cloud (Amazon EC2) and Microsoft Azure, both combined to a local machine. The choice of two different hybrid cloud environments aims at comparing the value added by the public cloud provider in testing different partitioning algorithms for agent-based systems.

The rest of the paper is organized as follows: Section 2 summarizes existing work related to the partitioning and distribution of applications on a hybrid cloud environment. Section 3 defines the proposed cost deployment model. And before concluding, section 4 describes the agent-based simulator developed as a case study, the description of the technical environment as well as the performed experiments and the discussion of the obtained results.

## 2. RELATED WORK

A careful literature study shows that hybrid cloud environments are more and more used to deploy and execute different types of distributed applications. In order to better summarize the existing works and the cost models adopted in each hybrid cloud deployment, we defined a set of five criteria to compare them:

Added to those already defined in (Juan-Verdejo and Baars 2013) to compare the partitioning of applications on a hybrid cloud infrastructure, we consider the metrics list and the exact description of the used hybrid environment, in terms of the public cloud and local premise. Therefore, the comparison, as shown in table 1, is performed according to the (1) focus, (2) the application type, (3) the granularity, (4) the used metrics and (5) the two-side hybrid cloud environment.

In (Hajjat Sun Sung Maltz Rao Sripanidkulchai and Tawarmalani 2010), the authors present the CloudWard Bound consisting in migrating enterprise services into a hybrid cloud environment by investigating the partition and reallocation at the server level. The approach considers the application performance requirements as well as the privacy restrictions to choose the candidate components for migration. To optimize the cost models associated with the use of hybrid clouds, CloudWard Bound implements a beneficial migration strategy based on considering the application characteristics including the workload amount, the storage capacity and transaction delays.

In (Van den Bossche Vanmechelen and Broeckhove 2010), the authors address an optimization problem consisting of deciding which workloads to outsource to what cloud provider when the private datacenters are overloaded. A binary integer program for hybrid cloud was proposed to formulate the workload scheduling problem during deployment while minimizing the total costs of the execution and meeting the users' requirements. Multiple metrics and constraints are defined in the binary integer program such as the deadline constraint, the execution time, the cloud providers, the monetary costs, etc. the authors define three scenarios including the use of public cloud only, public cloud with network costs and hybrid cloud. The solver implementing the optimization problem shows better results when it is used across public cloud provider but with hybrid cloud settings the performance of the solver decreases. The authors believe that enhancing the optimization technique with hybrid clouds presents an interesting research section.

In (Kaviani Wohlstadter and Lea 2012), the Manticore framework, which is a partitioning software service for deployment on a hybrid cloud, is presented. The authors identify two existing approaches found in the literature namely the request-based model and the static structure model. The first one consists in executing the partitioning software service request on the public cloud or on the local infrastructure. While, the second approach allows to partition a request between both of them. The existing approaches reveal some issues such as in flexibility in placement and cost-performance issues. To overcome the mentioned problems, the authors present a context sensitive model to automate the partitioning process that allows partitioning dynamically a request on both a public cloud and a local infrastructure while maintaining a distinguished transitive set of callers to each function execution. However Manticore addresses a low level of abstraction which is the code partitioning. The approach is dedicated to move code (functions) rather than components which is not ideal for enterprise applications which are developed by integrating different commercial tools and components. The authors propose a cost deployment model based on three metrics: execution time, communication overhead and monetary costs associated to the use of hybrid cloud infrastructure. The proposed model is used to compare the cost-performance of partitioning the application code on both the premise and public cloud (Amazon EC2) instead of using only one of them.

In (Fan Wang and Chang 2011), an agent-based migration framework for the migration of MapReduce jobs on a hybrid cloud is presented. Indeed, MapReduce consists into two tasks which are as follows: 1) The first phase is the map Job, which takes as input a set of data to be converted into tuples. 2) The reduce phase, takes as input the output of the previous phase and combines the received data into smaller tuples. The framework is based on agents to monitor system behavior, negotiate actions, manage resources and achieve automatic and intelligent service migration. The partitioning is based on three metrics including the job count, size of the job and estimated finish time. Unlike (Van den Bossche Vanmechelen and Broeckhove 2010) and (Hajjat Sun Sung Maltz Rao Sripanidkulchai and Tawarmalani 2010) the cost models are not taken into consideration.

In (Siddiqui Tahir Rehman Ali Rasooland Bloodsworth 2012), the Elastic-JADE (Java Agent Development Framework) platform using the Amazon EC2 resources, to scale up and down a local running agent based simulation is presented.

Table1: Research work related to the hybrid cloud partitioning approaches

| Research Work | Focus | Application Type | Granularity | Metrics | Hybrid cloud composition |
|---|---|---|---|---|---|
| (Hajjat Sun Sung Maltz Rao Sripanidkulchai and Tawarmalani 2010) | enterprise services migration | Enterprise Applications | Components | 1. Applications Number, 2. Components number per application, 3. Transaction number, 4. Size of a transaction, 5. Number of servers, 6. Monetary costs | Large campus Network combined to Windows Azure Cloud Platform |
| (Van den Bossche Vanmechelen and Broeckhove 2010) | Workload scheduling | simulation experiments, image and video rendering codes, highly parallel analysis codes | Jobs | 1.Applications Number, 2.Tasks number per application, 3.Cloud Provider, 4.Instance Type, 5.Execution time/task, 6.Deadline constraints, 7. Monetary costs. | Local infrastructure combined to multiple cloud provider (A, B,C) |
| (Fan Wang and Chang 2011) | Services migration | MapReduce | MapReduce jobs | 1.Jobs count, 2. Size of jobs, 3. Estimated finish time | Not mentioned |
| (Siddiqui Tahir Rehman Ali Rasooland Bloodsworth 2012) | Agents migration | Agent-based systems applied in Image processing | Agents | 1.CPU , 2. Memory | A local machine combined to Amazon EC2 instances |
| (Kaviani Wohlstadter and Lea 2012) | code entities partitioning | Web Applications | Code functions | 1. Execution Time 2. Communication time 3. Monetary costs | A local machine combined to a large Amazon EC2 instance |
| (Juan-Verdejo and Baars 2013) | BI applications migration strategies using a set of predefined criteria. | BI applications | Components | 1. Applications number 2. Components number per application. 3. Transaction frequency 4. Transaction delays 5. Cloud Costs | Not mentioned (no experiments) |

According to the authors, JADE allows the distribution of MAS, and combining a running local JADE platform to the cloud, can enhance the scalability of such systems. In this work, thresholds are defined to determine the right moment to scale up or down the application.

The proposed approach consists in using a monitoring system on both local and cloud resources, to keep track of the available resources and exchange messages. While the proposed solution enhances scalability, it is specific to the MAS developed using the JADE platform. However, the authors did not investigate the cost of migration, or the overhead of communication between local and cloud resources that occurs during simulation.

In (Juan-Verdejo and Baars 2013) a cloud migration framework for partially moving Business Intelligence (BI) applications in the hybrid cloud is described. The authors identify different criteria including functional and nonfunctional requirements related to BI applications to define multiple alternative migration strategies. The authors consider the use of the Analytic Hierarchy Process (AHP) which assigns weights to the criteria in order to assist the migration decision. The authors don't consider the cloud offering selection such as cloud advisory service, migration service, Cloud Environment Build and Management Service (EBMS), since such services affect the cost model as well as the QoS.

As it is described in the Table.1, the deployment of distributed applications on a hybrid cloud environment requires special considerations to reach cost-performance efficiency. Therefore different metrics were defined based on the characteristics of the applications, the used hybrid infrastructure and the policies of the user.

In the current work, we are interested in the deployment of distributed agent-based systems on a hybrid cloud infrastructure. An efficient partitioning approach as well as a suitable cloud provider are required to achieve both performance in terms of running large-scale distributed application and cost savings while using hybrid clouds. Similarly to the mentioned research work in Table.1, in order to accomplish the partitioning of an agent-based simulation on a hybrid cloud, a set of metrics should be taken into consideration.

Consequently, in the current work, based on the existing research work and the characteristics of an agent-based system, we define the relevant metrics to be considered while partitioning agent-based simulation applications on a hybrid cloud infrastructure: 1) the communicated data between the public cloud and the on premise infrastructure (in our case such data can be either the messages transfer between agents residing on different machines or the agents migrations between nodes); 2) execution time; 3) monetary costs related to the use of a public cloud provider. Each metric will be detailed and discussed in the next section.

## 3. COST DEPLOYMENT MODELLING

Usually, in order to evaluate a partitioning mechanism used with agent-based simulations, not in hybrid cloud environments, the following metrics are considered in several research works (Wang Lees Cai Zhou and Low 2009, Wang Lees and Cai 2012, Vigueras Lozano Orduña and Grimaldo 2010) :

- Communication cost: denotes the number of messages that are communicated between two agents residing in different machines.
- Migration cost: presents the number of migrations of the agents during simulation time. A migration denotes the move of an agent residing on a given machine to another different one.
- Execution Time

Also, the cost of a deployment on a hybrid cloud depends on the execution time as well as the size of the communicated data associated with the monetary costs offered by a cloud provider.

Combining these two points of view and resulting literature findings, in the current work, we matched the considered metrics in the partitioning agent-based simulation with those used to evaluate the cost of a hybrid cloud deployment in order to formulate the agent-based simulation-partitioning problem on a hybrid cloud. Therefore, we introduce the notations given in Table 2.

Before defining, the different costs associated with the deployment of a distributed agent-based simulation on the hybrid cloud, we made the following assumptions:

- All the communicated types of messages have the same size.
- All the agents have the same size.
- No load balancing during the simulation (static partitioning algorithms).
- The transfer of data is charged in one way (from cloud to local machine).

In the rest of this section, we define our cost model associated to the deployment of a distributed agent-based simulation on a hybrid cloud where we consider (1) The communication cost, which introduces the cost of sending one message from the public cloud to the local infrastructure; (2) The migration cost, which presents the cost of moving one agent from the public cloud to the promise; and (3) Execution Time.

Table2: Notations used to model agent-based systems deployment cost

| Symbols | Definitions |
| --- | --- |
| $a_i$ | agent i / (i=1,..,n) |
| $S_i$ | Size of an agent $a_i$ |
| $S_{msg}$ | Size of a message |
| $Cost_{send-one-msg}$ | Cost of sending one message from a cloud machine to a local machine |
| $Cost_{migrate-one-a_i}$ | Cost of migrating an agent from a cloud machine to a local machine |
| $T_{send-one-msg}$ | Time of sending one message from the cloud to the on premise machine |
| $T_{migrate-one-a_i}$ | Time of migrating one agent from the cloud to the on premise machine |
| $T_{Unit}$ | Time unit defined by the cloud provider to apply charges |
| $D_{Unit}$ | Data unit defined by the cloud provider to apply charges |
| $Cost_{Tunit}$ | Cost associated to $T_{Unit}$ |
| $Cost_{Dunit}$ | Cost associated to $D_{Unit}$ |
| $\beta$ | Bandwidth |
| $\lambda$ | Communication latency for a given cloud |

### 3.1. Communication Cost

The communication cost depends on the size of the transferred data as well as the time of communication. Thus, similarly to (Kaviani Wohlstadter and Lea 2012) we start by determining the time required to transfer data from a public cloud machine to premise machine. The first equation represents the time of transferring one message from an agent residing on the public cloud to another one located on the premise infrastructure.

$$T_{send-one-msg} = \frac{S_{msg}}{\beta} + \lambda \quad (1)$$

Where $S_{msg}$ indicates the size of a message communicated between two agents and β is the communication bandwidth between the cloud provider and the local infrastructure. The first part of the above equation $[\frac{S_{msg}}{\beta}]$ denotes the message transmission time between the cloud and the premise; and $\lambda$, as it is defined in several research work (Kaviani Wohlstadter and Lea 2012, Hajjat Sun Sung Maltz Rao Sripanidkulchai and Tawarmalani 2010, Van den Bossche Vanmechelen and Broeckhove 2010) as a constant representing the communication latency associated with the use of a given public cloud provider. Then, we associate the monetary costs offered by the cloud provider as follows:

$$\text{Cost}_{\text{send-one-msg}} = \frac{S_{msg}}{D_{Unit}} * \text{Cost}_{Dunit} + \mu * \frac{T_{\text{send-one-msg}}}{T_{Unit}} * \text{Cost}_{Tunit} \quad (2)$$

$$\mu \begin{cases} 0, \text{if the transmission delays are neglected} \\ 1, \text{Otherwise} \end{cases}$$

Where $D_{unit}$ is the unit of data defined by the cloud provider to apply the charges; also $\text{Cost}_{comunit}$ represents the cloud price associated to $D_{unit}$. The first part of the equation denotes the charges directly related to transferring one message between the public cloud and the local infrastructure. The second part introduces the cost associated to the time of sending one message. $T_{unit}$ presents the unit of time used by the cloud provider to apply charges and $\text{Cost}_{execunit}$ is the cloud cost associated to $T_{unit}$. Finally, we introduce, U, which is a binary variable that can take either zero if the time of sending one message is neglected or one in the opposite case. The second part of the equation, as it is defined, is more general and can model previous works: For (Kaviani Wohlstadter and Lea 2012), the time of communicating data is considered important and affects the overall cost of the deployment on a hybrid cloud environment, however in other research work such as in (Hajjat Sun Sung Maltz Rao Sripanidkulchai and Tawarmalani 2010, Van den Bossche Vanmechelen and Broeckhove 2010) it is neglected, only the size of the data does matter.

### 3.2. Migration Cost
Similarly to the communication cost, the migration cost depends on the size of the agent to be moved as well as on the time taken to process the migration in the hybrid cloud environment. Similarly, to the way used to calculate the communication cost, we determine the cost migration. Thus, we start by determining the time of moving one agent from a cloud machine to a local machine.

$$T_{\text{migrate\_one\_ai}} = \frac{S_i}{\beta} + \lambda \quad (3)$$

Where $S_i$ presents the size of the agent to be moved and β is the communication bandwidth between the cloud provider and the machine on premise. The first part of the equation $[\frac{S_i}{\beta}]$ denotes the transmission time; and $\lambda$, as it is defined above, is a constant representing the communication latency associated with the use of a given public cloud. To determine the migration cost we associate the monetary costs offered by a public cloud provider as follows:

$$\text{Cost}_{\text{migrate-one-ai}} = \frac{S_i}{D_{Unit}} * \text{Cost}_{Dunit} + \mu * \frac{T_{\text{migrate\_one\_ai}}}{T_{Unit}} * \text{Cost}_{Tunit} \quad (4)$$

$$\mu \begin{cases} 0, \text{if the transmission delays are neglected} \\ 1, \text{Otherwise} \end{cases}$$

The first part of the equation denotes the charges directly related to the migration of one agent from the public cloud to the local machines; and the second part introduces the cost that can be added due to the migration time. This factor either it is neglected when U is equal to zero or considered in the overall migration costs. The objective from adding the second part in both the communication and migration costs is to evaluate the effect of transmission time on the overall deployment cost.

### 3.3. Execution Time Cost
The computational cost depends mainly on the characteristics of the used cloud virtual machines (VM), the execution time and the load running on the VM. If we assume that we are using the public cloud resources during the whole simulation (from the beginning until the end of the simulation) then the time of using the cloud machine will be equal to the overall execution time of one simulation.

$$\text{Cost}_{exec} = \frac{T_{exec}}{T_{unit}} * \text{Cost}_{exe\_Unit} \quad (5)$$

### 4. CASE STUDY AND EXPERIMENTATION
In this section, we present the description of the used agent-based simulator, the description of the technical environment as well as the experimental results of conducting the distribution of an implemented agent-based simulator on two different hybrid cloud providers using the defined cost model. The distribution is performed using two partitioning algorithms, which are the K-means and the static environment partition.

### 4.1. Description of the agent-based model
The model consists of N agents moving within a spatial environment represented as a grid and interacts between each other. According to the Vowels model (Demazeau 1995), we highlight the environment proprieties, the agents' characteristics as well as the interaction aspect.

- Environment: In the current work, we are interested in the conceptual level of MAS; therefore, we consider only the application environment where the agents and objects are embedded. In our agent model the environment represents the geographical area in which the agents move. It is characterized by its height and width.
- Agents: Our model contains two types of agents including rescuers and victims. The agents are initially assigned to random positions. 1) The rescuers: present the leaders who move to the nearest exit; 2) The victims: follow the nearest rescuer in term of spatial distance to reach the emergency exit. The agents can move, communicate and migrate from one node to another.
- Interactions: Agents in a MAS should be able to interact and understand each other in order to coordinate their actions and possibly cooperate. Therefore, similarly to (Wang Lees Cai Zhou and Low 2009, Wang Lees and Cai 2012) ,we define an Area of Interest (AoI). Agents can communicate and change messages when their AOI overlaps.

To summarize, the test case considered in this current work consists of a set of agents moving within an environment (100 cells *100 cells). We consider an emergency situation where the agents are organized into two different categories: victims and rescuers. The rescuers present the leaders who move to the exits and the victims follow the nearest rescuer in terms of spatial distance. We used the Jade agent platform [20] to develop the simulator described above.

### 4.2. Description of the used partitioning algorithms

Based on the existing research work (Solar Suppi and.Luque 2011, Solar Suppi and.Luque 2012, Vigueras Lozano Orduña and Grimaldo 2010), the partitioning mechanisms can be classified into two different families:

- Clusters-based approaches in which the agents are clustered based on given criteria then assigned to different nodes.
- Grid-based approaches, called also region-based approach consists of decomposing the space component (environment) into multiple portions. Each portion together with the set of agents residing in it are assigned to a node.

Both cluster and grid approaches can be either static or dynamic. Indeed, the static strategies distribute the agents without considering the load balancing. However, the dynamic mechanisms are used to load balance the partitions during simulation time. In the current work, we consider two partitioning algorithms, which are the K-means and the static environment partition.

The K-means clustering algorithm referred also as the Lloyd's algorithm (Lloyd 1982), was elaborated by Forgy (Forgy 1965) and McQueen (McQueen 1967).

Indeed, the K-means is a cluster-based algorithm consisting of grouping objects into clusters based on a given set of attributes. Indeed, K presents an initial input which denotes the number of the clusters. Each cluster has a centroid which can be determinate by calculating the central mass of all objects. The algorithm starts by initializing the centroids and assign objects to the closest centroid. The process repeats till there is no changes in the clusters.

In this current work, we used the agents' spatial locations as an attribute to cluster them into groups. Each agent is assigned to the cluster with the nearest centroid. We consider a number of clusters equal to four (k=4) including three partitions running on the local machine and a single one on the public cloud machine.

The other method is a grid-based approach represented by the static environment partition. It consists of decomposing the spatial component into regions or cells and assign each partition with the agents residing on it to a single node.

Using this method it is possible to decompose the environment in a number of ways: in a regular grid, in vertical/horizontal strips or in an irregular grid (Wang Lees Cai Zhou and Low 2009).

In order to compare its performance with the above cluster-based algorithm, we choose to decompose the application environment into four partitions using the regular grid division. Similarly to the K-means, including three partitions running on the local machine and a single one on the public cloud machine. Indeed, the grid division is described in (Wang Lees Cai Zhou and Low 2009) as the appropriate choice to minimize the communication overhead between partitions.

### 4.3. Description of the technical environment

We performed experiments to compare the performance of K-means as well as the static environment partition on different hybrid clouds environments. As a local machine, we used a 2.10 GHz i3-2310M CPU with 4.0 Go memory. For the public cloud machines, we used on Amazon EC2, an m1.large machine from west Europe with 4 EC2 compute units and 7.5 Go memory. For Microsoft Azure, we used the equivalent to m1.large EC2 instance, which is A3 large type with 4 CPU cores and 7 Go memory. In both cases, using either Amazon EC2 instance or Microsoft Azure, the local and cloud machines were connected with a data link of 100 Mb/sec. For the latency, we measured 124.1 milliseconds for Amazon EC2 and 163.6 milliseconds for Microsoft Azure.

For all the experiments, the Amazon EC2 instance costs 0.19$ per hour and the A3large on Microsoft Azure is charged for 0.324$ per hour. For the data transfer charges it was the same for both clouds 0.12$ from the cloud machine to the on premise one and 0$ in the opposite case.

### 4.4. Evaluation

We evaluate both algorithms including the K-means algorithm and the static environment partitioning method using our proposed deployment cost model.

We conducted our experiments using different configurations summarized in Table.3.

The results of our experiments are presented as the mean of 10 runs for each configuration. As it is shown in Table.3, initially, the partitions have an equal number of agents including one rescuer in each partition. However, since both of the used algorithms are static, each simulation ends up with an extremely unloaded partitions.

Table 3: Configurations' characteristics

| Config. | Total Number of Agents | Number of Victims | Number of rescuers | Number of agents on local machine | Number of agents on the cloud |
|---|---|---|---|---|---|
| Config.1 | 200 | 196 | 4 | 150 | 50 |
| Config.2 | 400 | 396 | 4 | 300 | 100 |
| Cnfig.3 | 600 | 596 | 4 | 450 | 150 |
| Config.4 | 800 | 796 | 4 | 600 | 200 |
| Config.5 | 1000 | 996 | 4 | 750 | 250 |

In fact, in the case of using both the K-means and the static environment partition, a grouped movement of the agents is more efficient than considering the individual movement pattern.

Therefore in the presented results, we consider only the case where the agents are initially clustered and moving into groups to follow the nearest rescuer.

### 4.4.1. Costs Evaluation with neglected transmission Time (U=0)

Indeed, in the proposed cost deployment model, within both defined equations to determine the migration as well as the communication costs we define a binary variable U, which take 0 if the transmission time is neglected or "1" in the opposite case.

The mentioned variable U is defined in the order to determine the effect of the transmission time on the overall costs. In this section, we present the costs evaluations without adding the transmission delays.

The Fig.1 presents the communication costs (message transfer) associated to the execution of both partitioning algorithms (K-means and Static environment partition) on both used hybrid clouds environments. As a first result the cluster-based algorithm outperforms the grid-based method on both hybrid cloud environments in term of the communicated messages between the cloud machine and the local one. Also, the communication costs increase with the used different configurations (agents number).

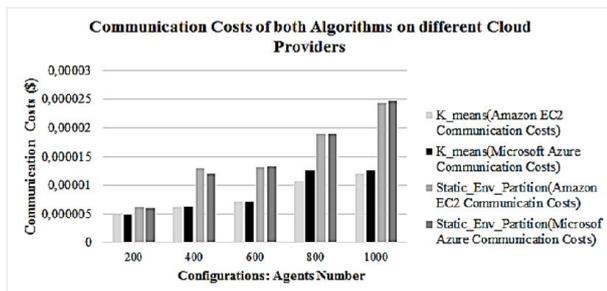

Figure 1: Communication Costs of both Algorithms on different hybrid cloud settings.

Based on previous experimental results in (Wang Lees Cai Zhou and Low 2009, Wang Lees Cai 2012, Labba Bellamine and Dugdale 2015), considering the movement pattern of agents is of great importance for the partitioning mechanism of a distributed agent-based system.

The Fig.2 represents the migration Costs of both used partitioning algorithms on different hybrid clouds settings. Similarly to the communication costs results, the K-means outperforms the grid-based method represented by the static environment partition. As it is shown in Fig.2, the overall costs of migrations increases with the used number of agents and the costs associated to static environment partition are much higher than those associated to K-means on both hybrid clouds settings either using Amazon EC2 or Microsoft Azure.

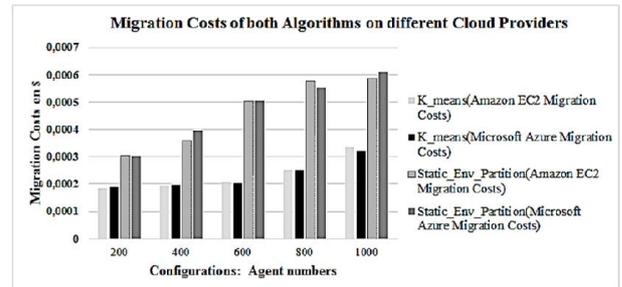

Figure 2: Migration Costs of both Algorithms on different Cloud Providers.

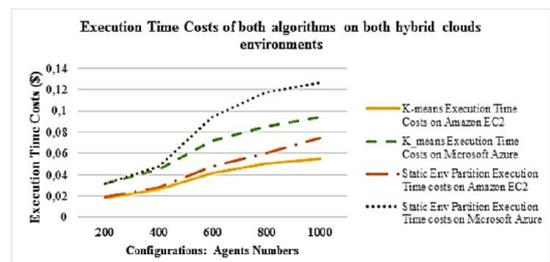

Figure 3: Execution Time Costs

The Fig.3, presents the execution time costs associated to both partitioning algorithms executed on both hybrid clouds (the local machine combined to either Amazon EC2 or Microsoft Azure).

As it is shown in Fig.3 the execution of K-means on

Amazon EC2 represents better results in term of execution costs compared to the static environment partition( on both hybrid clouds environments) and even to K-means tested on Microsoft Azure. Also, with both hybrid cloud settings K-means outperforms the static environment partition in term of execution time. To summarize, based on the results of the experiments shown respectively in Fig 1, Fig 2 and Fig 3 the K-means algorithm shows better results in terms of {migration, communication and execution time} costs in both cases of using hybrid cloud environment with different settings (Amazon EC2, Microsoft Azure).

For both migration and communication aspects, the overall costs of running the K-means (same for the static environment partitioning) using both the public clouds Amazon EC2 and Microsoft Azure shows close results. Thus, we can explain these results, by the fact that both of the public clouds charges the data transfer from cloud machines to on premise ones using the same prices.

However, for the execution time costs, shown in Fig.3 is more dependent to the used public cloud provider. Both of algorithms including K-means and static environment partition show better results with Amazon EC2. The increase of execution costs with Microsoft Azure can be returned to the high prices suggested by the latter one to use cloud VMs compared to Amazon EC2.

### 4.4.2. Costs Evaluation including transmission delays (U=1)

In the rest of this section, we present the effect of the transmission delays on the overall costs for the K-means algorithm executed on both used hybrid clouds environments.

In one hand, Fig.4 represents the communication costs including the transmission delays for K-means on both hybrid clouds environments. When the binary variable U is equal to "1", the overall costs of communication on both hybrid clouds settings are much expensive than the case where U is equal to zero.

Also, as it is shown in the Fig.4, the costs are much important with Microsoft Azure which can be returned to the high prices suggested by this latter one.

In the other hand, Fig.5 describes the migration costs including the transmission delay for K-means on both hybrid clouds environments. Similarly to the results presented with the communication costs, the transmission delay has an impact on the overall migration costs. This impact is higher with the use of Microsoft Azure compared to Amazon EC2.

To summarize, as it is shown respectively in both Fig.4 and Fig.5, the transmission delay has an important effect on the overall costs for communication as well as migration for both algorithms executed on the different hybrid clouds settings.

However, this difference is more important with Microsoft Azure due the high prices suggested by this latter one compared to Amazon EC2.

The costs depend on three aspects: 1) the size of the transferred data between the public cloud and on the local machine. Once the overall size of data is important, the costs are important too; 2) the transmission delays are related mostly to the state of the network and the size of the transferred data; 3) the used public cloud provider. Indeed, the hybrid cloud settings is important and should be considered to minimize the overall costs of deployment.

Therefore for distributed agent-based systems, choosing the suitable partitioning method has a huge effect on the overall costs of hybrid cloud deployment. A suitable partitioning approach has to minimize the migration and communication overheads as well as the execution time. In addition, the choice of a public cloud provider is much important as the partitioning method, since each provider has its own prices and manner of charging costs.

To summarize, the deployment challenge of a distributed agent-based model on a hybrid cloud infrastructure returns to a decision making process permitting to select the best partitioning algorithm and the suitable hybrid cloud composition.

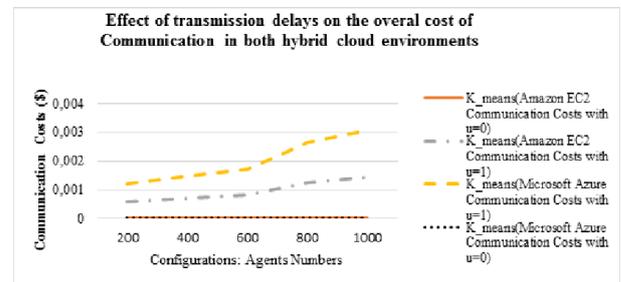

Figure 4: Communication Costs Including the delay costs for K-means on different cloud providers

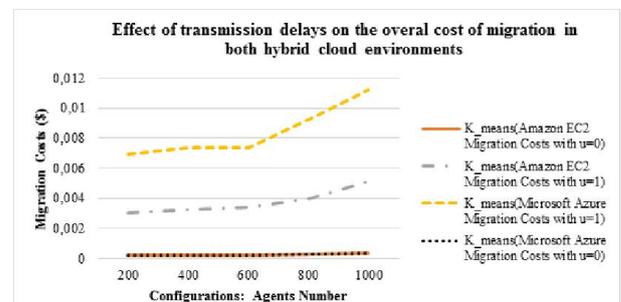

Figure 5: Migration Costs Including the delay costs for K-means on different cloud providers

### 5. CONCLUSION

To enhance the distribution of agent-based systems both an efficient partitioning method and a reliable/powerful infrastructure are required.

In this paper we evaluated the cost-performance of both cluster and grid based partitioning methods on two different hybrid cloud environments.

The main contributions of this paper are as follows:
- A cost deployment model dedicated to agent-based systems is proposed in order to evaluate the performance of partitioning methods and the suitability of the hybrid cloud environment. Indeed the model takes into consideration mainly the communication overhead between the partitions, the number of migrations of agents, the execution time, the latency, the bandwidth and the costs associated to the use of the public cloud.
- Comparison of both partitioning algorithms including k-means and the static environment partition executed on two different hybrid cloud environments. The experimental results show that k-means algorithm outperform the static environment partition on both used hybrid infrastructure. Also we discovered that the transmission delays has an impact on the overall costs.

Therefore , the deployment challenge of a distributed agent-based model on a hybrid cloud infrastructure returns to a decision making process permitting to select the best partitioning algorithm and the suitable hybrid cloud composition. As a future work, we aim at automating this decision making process using a set of relevant criteria  as well as the distribution and partitioning of an agent-based model on a hybrid cloud infrastructure.